# Critical optical coupling between a GaAs disk and a nanowaveguide suspended on the chip

C. Baker[1], C. Belacel[2], A. Andronico[1], P. Senellart[2], A. Lemaitre[2], E. Galopin[2], S. Ducci[1], G. Leo[1], I. Favero[1] *

[1] *Université Paris Diderot, Sorbonne Paris Cité, Laboratoire Matériaux et Phénomènes Quantiques, CNRS-UMR 7162, 10 rue Alice Domon et Léonie Duquet, 75013 Paris, France*

[2]*Laboratoire de Photonique et Nanostructures, CNRS, Route de Nozay, 91460 Marcoussis, France*

* E-mail: ivan.favero@univ-paris-diderot.fr

We report on an integrated GaAs disk/waveguide system. A millimeter-long waveguide is suspended and tapered on the chip over a length of 25 µm to evanescently couple to high Q optical whispering gallery modes of a GaAs disk. The critical coupling regime is obtained both by varying the disk/guide gap distance and the width of the suspended nanoscale taper. The experimental results are in good agreement with predictions from the coupled mode theory.

Whispering Gallery Mode (WGM) Gallium Arsenide (GaAs) optical cavities combine the advantageous optical properties of GaAs with small mode volumes and high quality factors Q, enabling a boost of light-matter interaction. Thanks to these assets, they have become popular in different contexts: for quantum electrodynamics experiments[1-3], for the realization of lasers[4] or non-linear optics devices[5], and now in optomechanics applications[6-8]. The best Q factors reported so far on GaAs WGM optical cavities are of a few $10^5$, measured by means of evanescent coupling to an optical fiber taper[6,8,9]. However, even if sufficiently robust for laboratory experiments, fiber tapers suffer from a poor mechanical stability and are affected by a rapid degradation in standard humid environments[10]. An integrated optics approach, where access to the cavity is provided directly on the chip by a bus waveguide, naturally solves these issues and increases the range of operation of GaAs WGM cavities. For example, a GaAs waveguide/disk structure was developed for quantum electrodynamics experiments[11]. In this structure, optical probing through a narrow-band grating coupler allowed to observe evanescent coupling in the under-coupled regime. However, for many applications, obtaining the critical coupling is crucial[12]. This applies notably in optomechanics[13-15] where the number of photons in the cavity needs to be maximized, both for cavity cooling and for optomechanical self-oscillation[16-18].

Here we investigate a GaAs waveguide/disk integrated structure but adopt a direct injection at the cleaved facet of the waveguide, with a two-fold advantage: first, simple conventional optics can be used for waveguide coupling; second, the coupling is wavelength independent and allows broadband (here 100nm) spectroscopy of the system. We observe the evanescent coupling of a GaAs disk to a nano-tapered waveguide suspended in the disk vicinity. By varying the disk/guide gap distance or the guide taper width, the overlap between disk and guide optical modes is adjusted and the important critical coupling regime is reached in a controlled manner. Our experimental results are in excellent agreement with coupled mode theory expectations[19,20].

We employ a semi-insulating commercial GaAs substrate, on which we grow by molecular beam epitaxy (MBE) a GaAs 500 nm buffer layer, a 1.8 µm thick $Al_{0.8}Ga_{0.2}As$ sacrificial layer and finally a 200 nm GaAs top layer. The guides and disks are defined by e-beam lithography, using a negative resist. We draw straight optical waveguides that are 5 microns wide and 2 mm long, extending up to the sample facets (Fig. 1a). The central region of each guide is tapered in the vicinity of the target disk, to allow evanescent coupling (Inset of Fig. 1a). The nominal disk diameter is 7.4 µm. The two other disks seen in the inset of Fig. 1a serve as fabrication witnesses and are not considered in what follows. In a first rapid (6 s) wet etch by a non-selective HBr-based solution, we separate the guide from the disk and define the lateral boundaries of the waveguide structure, etching away the top 200nm of GaAs to attain the sacrificial layer. In a second etch step, we use dilute (2%) Hydrofluoric Acid (HF) to selectively under-etch the AlGaAs layer, with several consequences. First, this results in locally suspending both the disk and the waveguide (see Fig 1b) with a gap distance of a few hundreds of nanometers between the two (see Fig 1c). Second, the lateral edges of the 5-micron wide waveguide are also under-etched on the whole length of the sample, resulting in a rail-like guiding structure (Fig. 1d). In this structure the transverse electric (TE) in-plane polarized fundamental mode is confined in the upper 200 nm GaAs layer by the semiconductor/air interfaces, resulting in a limited mode leaking into the substrate. Conversely, Beam Propagation Method (BPM) simulations predict that the out-of-plane polarized (TM) fundamental mode leaks more importantly into the substrate, leading to important optical losses of the rail-guide, which are confirmed experimentally. As a result, our waveguides efficiently select the TE polarization. Injection into the waveguide is performed at the facet, where the rail-like guiding structure is cleaved with a straight angle (Fig. 1d). The waveguide taper profile is a compromise between geometric simplicity and good adiabaticity: we have chosen a 3-part linear profile, with respective lengths of 10, 5 and 10 µm. The width of the central part of the taper varies from guide to guide between 200 and 380 nm, generally ensuring a single TE guided mode in this zone. For the chosen profile, BPM calculations predict a transmission of the fundamental mode through the taper always superior to 90 %.

Continuous wave optical spectroscopy of the disk/waveguide system is performed in the 1300-1400 nm band. The beam of a tunable external cavity diode laser (linewidth < 1 MHz) is focused onto the input facet of the guide using a micro-lensed fiber with a waist diameter of 2.3 µm. The TE polarization is selected by a fiber-polarization-control. Light is collected by a microscope objective at the output facet of the guide and sent onto an InGaAs PIN-photodetector. All experiments are carried out under ambient conditions. Fig 2 shows the transmission of a guide evanescently coupled to a disk as a function of the laser wavelength between 1340 and 1420 nm. The transmission has been normalized here, to compensate for a progressive reduction of the guide transmission when scanning the laser towards longer wavelengths. This reduction is due to a residual leaking of the guided mode into the substrate and is well reproduced by BPM simulations. It can be reduced by further under-etching the lateral edges of the rail waveguide, in order to better confine the guided mode in the top GaAs layer. Several resonances appear in the spectrum of Fig 2, with a transmission contrast up to 80%. They correspond to WGM optical resonances of the disk, whose best loaded Q factors reach 55000, as shown in the inset of Fig 2. This corresponds to intrinsic optical Q factors in the high $10^4$ range. We note the appearance of pronounced Fano features involving broad and fine optical resonances. Similar profiles were already observed in the spectrum of a silica WGM toroid cavity coupled to a fiber taper, and can be attributed to interference between different modes of the cavity[21]. In our system, axis-symmetric Finite Element Methods calculations[22] show that low-Q WGMs correspond to large radial $p$ numbers (up to $p=7$ for the present disks), while high Qs are observed on smaller $p$ modes. Our spectra show reproducibly Fano interferences between large radial number $p$ and small-$p$ modes of the GaAs disks. The full labeling of the observed WGMs with azimuthal and radial numbers[23] $m$ and $p$ will be discussed in detail elsewhere.

In the present article we focus on the possibility of finely controlling the disk/waveguide evanescent coupling in our integrated system. This coupling can be tuned in several manners, to reach the critical regime where intrinsic losses of the cavity equal losses due to the coupling to the guide. Our first strategy if to vary the gap distance between disk and fiber to adjust the overlap of the two coupled optical modes[12,8]. Fig. 3a shows a series of transmission spectra obtained by varying the disk/guide gap distance g but keeping a constant taper width of 320 nm. From spectra 1 to 8, the gap distance is reduced in 25 nm steps from 350 to 190 nm. The gap distances are measured in a Scanning Electron Microscope (SEM) with an uncertainty of ± 7 nm, and agree within 5 % with nominal values. At small gap distances (most notably in the last spectra 5 to 8 in Fig 3a), several resonances approach zero transmission, reflecting critical coupling between the corresponding disk WGMs and the waveguide mode.

Let us now follow in detail the behavior of the resonance that we have marked with a star in the series of spectra of Fig. 3a. Fig 3b and 3c show the on-resonance normalized transmission $T_{on}$ and the linewidth $\delta\lambda$ of the selected resonance as a function of the guide/disk gap distance g. As g decreases, the overlap between the WGM and the guide mode increases, increasing the evanescent coupling. In Fig. 3b, $T_{on}$ is progressively decreased when g decreases, before reaching zero when the critical coupling is obtained for a gap distance g of 250 nm. Smaller gap distances make the system enter the over-coupled regime, where coupling losses overcome intrinsic losses of the disk. In Fig. 3c, we see that the linewidth $\delta\lambda$ is progressively enlarged as the evanescent coupling is increased, reflecting supplementary losses of the WGM due to its coupling to the guide.

Coupled mode theory (CMT) can be used to describe the evanescent coupling of the related WGM and its transition from the under to over-coupled regime, as the gap distance is reduced. A heuristic CMT[19] approach leads to remarkably simple expressions for the taper/disk on-resonance transmission $T_{on}=[(1-\gamma_e/\gamma_i)/(1+\gamma_e/\gamma_i)]^2$ and for the resonance linewidth $\delta\lambda=(\lambda_0/Q_{int})(1+\gamma_e/\gamma_i)$, where $\gamma_e$ is the extrinsic WGM-to-guide coupling rate, $\gamma_i = (2\pi c/\lambda_0)/Q_{int}$ is the intrinsic WGM loss rate, $\lambda_0$ the resonance wavelength and $Q_{int}$ the related intrinsic Q factor. These expressions were used to describe fiber taper evanescent coupling experiments with $\gamma_e$ as a parameter varying exponentially with the gap distance g [8,24,25]. In this approach, $\gamma_e$ reads $\gamma_e(g)=\gamma_e(0)\exp(-\eta g)$ with $\eta$ the inverse of the decay length of the evanescent coupling. In Fig. 3b, the solid line is a fit using this exponential approximation and the above formula for $T_{on}$, and taking $\gamma_e(0)/\gamma_i$ and $\eta$ as adjustable parameters. The best agreement with experimental data is obtained for $\gamma_e(0)/\gamma_i=12.6$ and $\eta=1/97\text{nm}^{-1}$. This value of $\eta$ represents the spatial extent of the evanescent coupling between the disk and guide modes. In Fig 3c, the experimental data are also fitted by a solid line corresponding to the above formula for $\delta\lambda$, for exactly the same fit parameters as above and where we use an average $Q_{int}$ of 8100. However, the nanofabrication tolerances result in a slightly different disk for each gap distance, and the level of intrinsic WGM losses represented by $Q_{int}$ cannot be considered as a constant. In Fig. 3c, we capture this variability by bounding the results with two dashed curves. These two curves are obtained with the formula for $\delta\lambda$ with the same value of $\gamma_e(0)/\gamma_i$ and for two bounding values of $Q_{int}$ = 7040 and 10280. These bounding values of $Q_{int}$ can then be injected back into the formula for $T_{on}$, with the same value of $\eta$, and with an average value of $\gamma_e(0)$ extracted from $\gamma_e(0)/\gamma_i=12.6$ and from the average $Q_{int}$ of 8100. This leads to the two dashed curves of Fig. 3b, which correctly bound the experimental results. To summarize, our experimental data are well fitted by taking into account a variability of ± 20 % of $Q_{int}$. We have obtained a similar level of agreement on other optical resonances of the same spectrum, with the noteworthy feature that the obtained $\eta$ is about constant (within a few percent) for the different resonances. This is consistent with numerical simulations, which predict nearly the same evanescent spatial decay for the different WGMs of a given disk.

We finally present a second independent manner of obtaining the critical coupling regime. In our integrated disk/guide device, we can directly modify the evanescent tail of the waveguide mode by a change of the guide width in the lithographic design. This enables a complementary engineering of the evanescent coupling and circumvents the difficulty of having to bring the disk and guide arbitrarily close to one another (indeed our isotropic wet etch prohibits us from reaching perfectly controlled gap distances below 200nm). Fig. 4 shows a series of transmission spectra obtained by varying the guide taper width, for a constant gap distance of 220 nm. From spectrum 1 to 6, the taper width varies in steps of 20 nm, from 380 to 280 nm respectively. At intermediate taper widths (spectra 3 and 4 of the series), several resonances approach zero transmission, reflecting critical coupling between the corresponding WGMs and the waveguide mode.

In summary, we have developed an integrated GaAs disk/waveguide system, where the suspended nanoscale tapered part of a GaAs guide couples evanescently to WGMs of a disk cavity. We observed the critical coupling regime and tested experimentally predictions of CMT. Our system offers straightforward optical access to on-chip micron-sized mode volume GaAs cavities of high Q. Perspectives include integrated quantum optomechanics experiments[26-29], where the strong optomechanical coupling of GaAs disk resonators[6,7] will be directly beneficial. The GaAs environment would additionally allow for coupling active optical elements, possibly electrically pumped, to optomechanical functionalities.

This work was supported by C-Nano Ile de France and the French ANR.

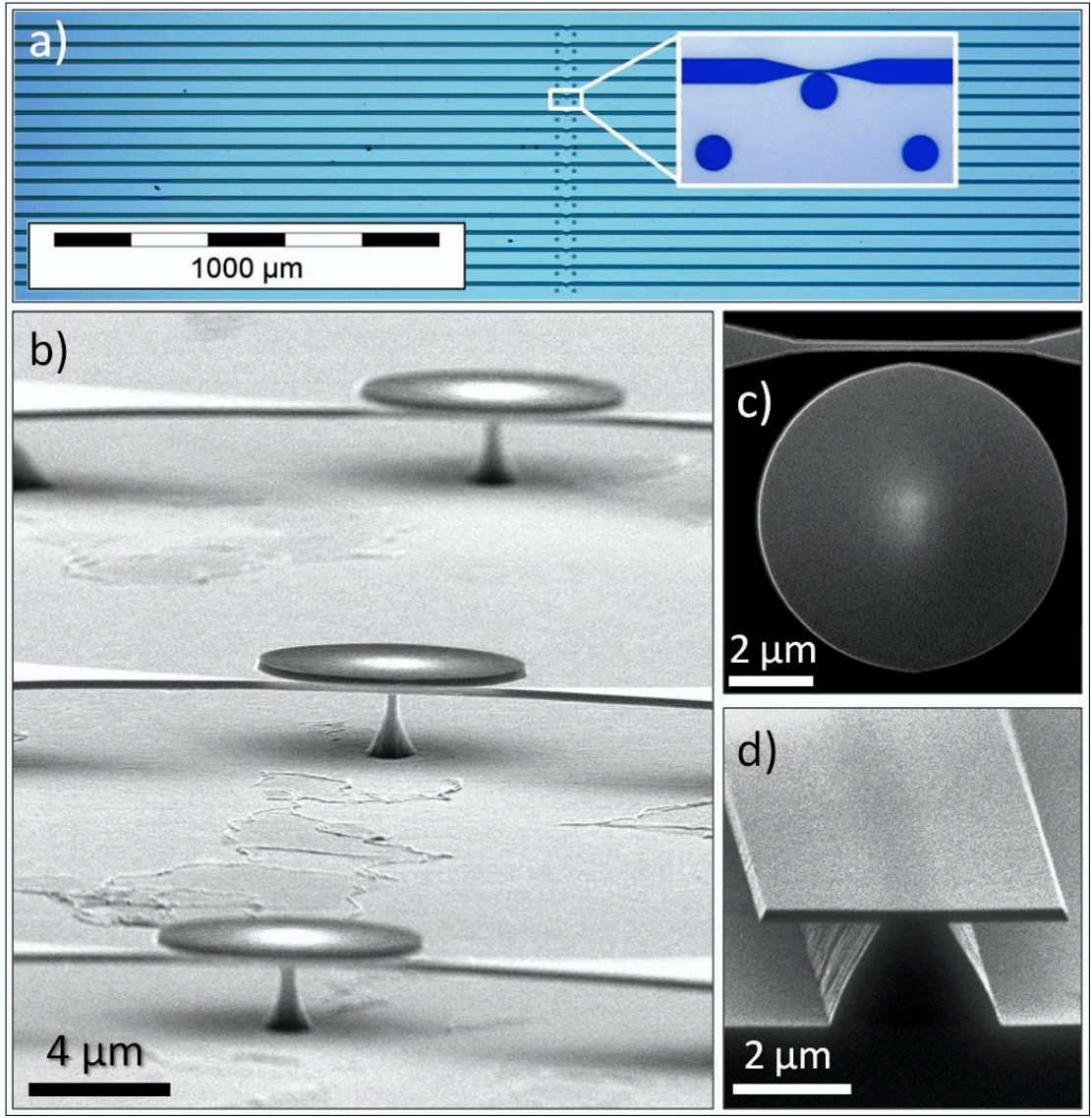

FIG. 1. GaAs waveguide/disk integrated device. a) Optical top view of the complete sample, containing 16 guide/disk systems. The inset is a close-up of the tapered part of the guide in the vicinity of the disk. b) SEM side-view of 3 disk/waveguide systems. c) SEM top view of the central part of the tapered waveguide next to a disk. d) SEM side view of the cleaved input facet of the rail-guide.

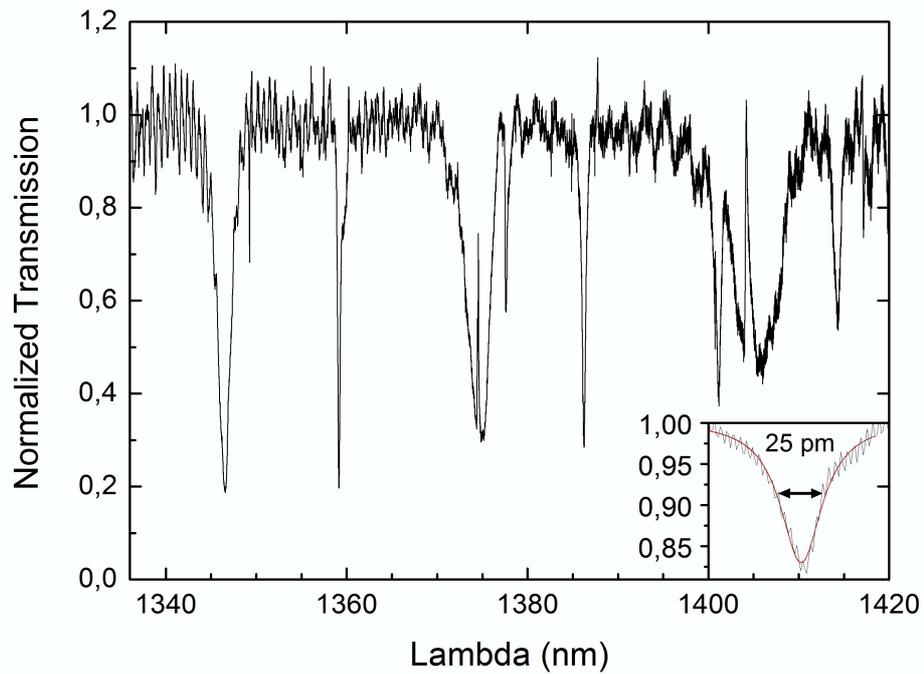

FIG. 2. Optical transmission spectrum of a suspended GaAs tapered waveguide coupled to a 7.4 μm diameter GaAs disk. The inset shows a fine optical resonance at λ=1370 nm with a loaded quality factor of 55000. The small amplitude oscillations in the base transmission correspond to Fabry-Perot fringes of the waveguide and are not noisy in nature.

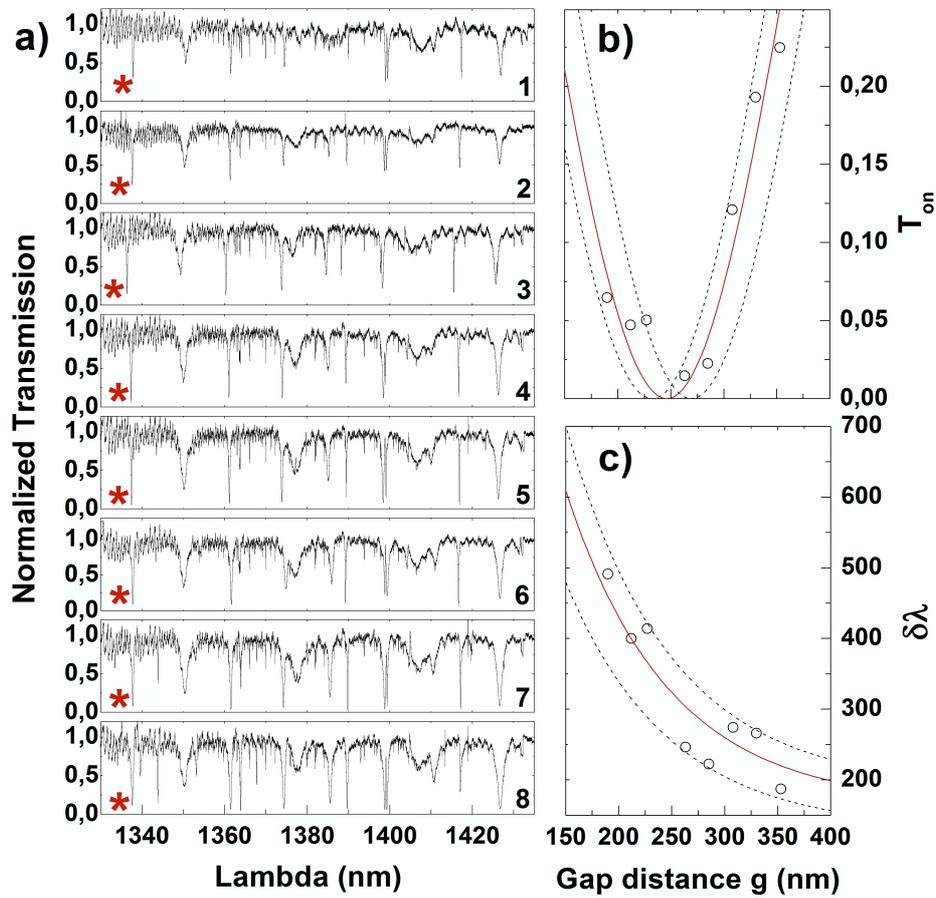

FIG. 3. Evanescent coupling for varying disk/guide gap distances g. a) Optical transmission spectra of suspended waveguides coupled to GaAs disks. From spectrum 1 to 8, g diminishes from 350 to 190 nm in steps of 25 nm. b) On resonance transmission $T_{on}$ as a function of gap distance g, for the resonance marked with a star in a). The open symbol size represents our experimental uncertainty in g and $T_{on}$. c) Width (in pm) of the same resonance as a function of g.

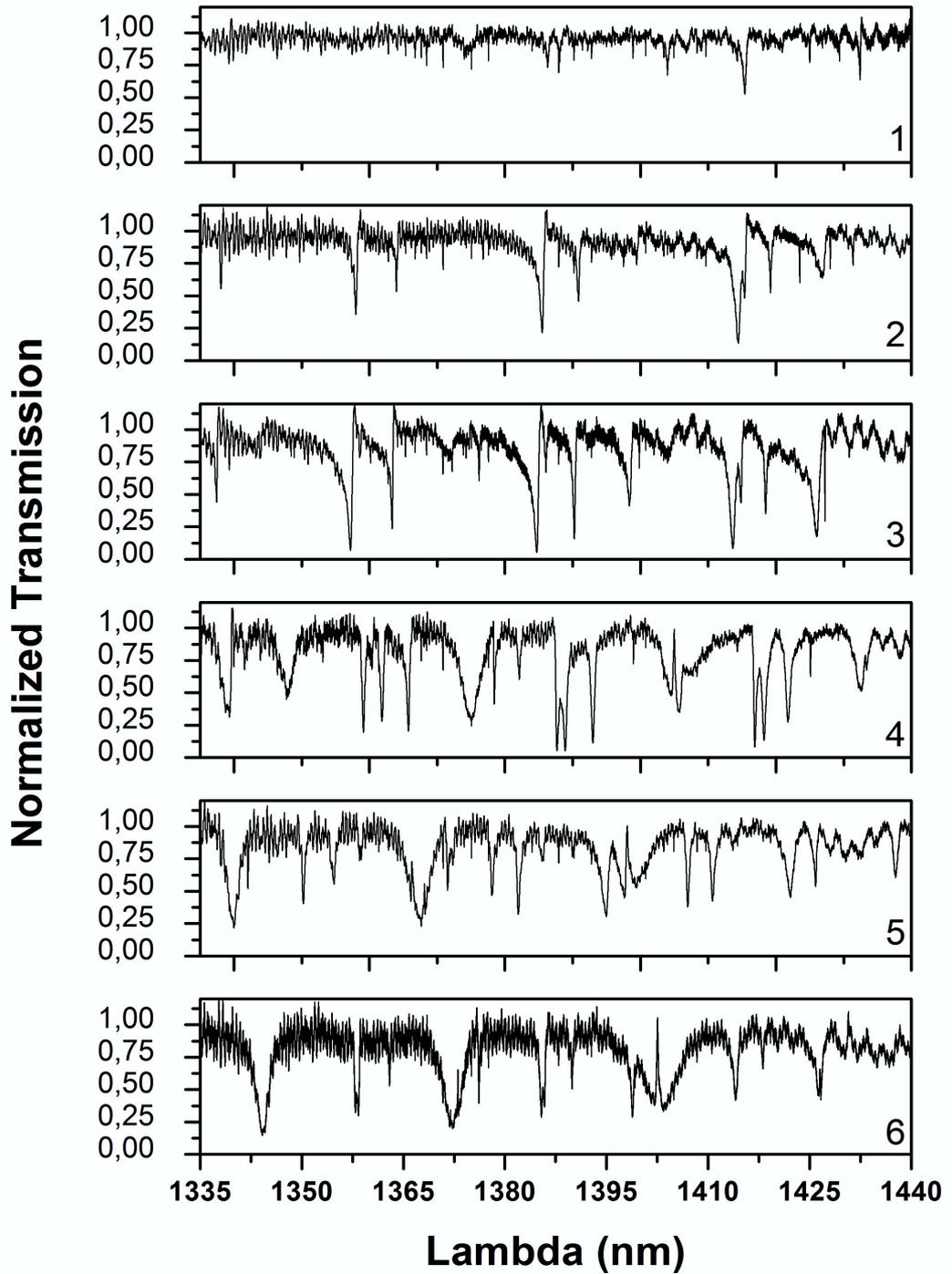

FIG. 4. Optical transmission spectra of suspended waveguides coupled to GaAs disks, as a function of the guide taper width. From spectra 1 to 6, the taper width diminishes from 380 to 280 nm in 20 nm steps.